\documentclass[aps,prl,twocolumn,groupedaddress]{revtex4}
\usepackage{amsmath}
\usepackage{amssymb}
\usepackage{graphicx}

\begin{document}

\title{Spins coupled to a Spin Bath: From Integrability to Chaos}

\author{John Schliemann}

\affiliation{Institute for Theoretical Physics, University of 
Regensburg, D-93040 Regensburg, Germany}

\date{\today}

\begin{abstract}
Motivated by the hyperfine interaction of electron spins with
surrounding nuclei,
we investigate systems of central spins coupled to a bath of noninteracting
spins in the framework of random matrix theory. With increasing number of 
central spins a transition from Poissonian statistics to the Gaussian 
orthogonal ensemble occurs which can be described by a generalized Brody
distribution. These observations are unaltered upon applying an external 
magnetic field. In the transition region, the classical counterparts of the
models studied have mixed phase space.
\end{abstract}

\maketitle


Spins coupled to a bath of other spin degrees of freedom occur in a
variety of nanostructures including semiconductor quantum dots
\cite{Petta05,Koppens06,Hanson07,Braun05}, carbon nanotube
quantum dots \cite{Churchill08}, phosphorus donors in silicon \cite{Abe04},
nitrogen vacency centers in diamond \cite{Jelezko04,Childress06,Hanson08},
and molecular magnets \cite{Ardavan07}. A large portion of
the presently very high both experimental and theoretical interest
in such systems is due to proposals to utilize such structures for
quantum information processing \cite{Loss98,Kane98,Leuenberger01}. 
Here the central spins play the role of the
qubit whereas the surrounding bath spins act as an decohering environment.
In the present letter we investigate very basic properties of such
so-called central spin systems in terms of spectral statistics
and random matrix theory \cite{Guhr98}.

The generic Hamiltonian is given by
\begin{equation}
{\cal H}=\sum_{\alpha=1}^{N_{c}}\vec S_{\alpha}\cdot\sum_{i=1}^{N}A_{i}^{(\alpha)}\vec I_{i}
\label{genmod}
\end{equation}
describing $N_{c}$ central spins $\vec S_{\alpha}$ coupled to $N$ bath spins
$\vec I_{i}$, typically $N\gg N_{c}$. Here we take all spins to be dimensionless
quantum variables such that the coupling constants $A_{i}^{(\alpha)}$ have 
dimension of energy. A paradigmatic example is given by, say, a single
spin of a conduction-band electron residing in a semiconductor quantum dot
and being coupled via hyperfine contact interaction to the
bath of surrounding nuclear spins. In a very typical material like
gallium arsenide all nuclei have a spin of $I=3/2$ whereas in other systems
like indium arsenide even spins of length $I=9/2$ occur. In fact, this
hyperfine interaction with surrounding nuclei has been identified to be
the limiting factor regarding coherent dynamics of electron spin qubits
\cite{Petta05,Koppens06,Hanson07,Khaetskii02}. 
In the above example the hyperfine coupling constants $A_{i}^{(\alpha)}$ are
proportional to the square modulus of the electronic wave function at the 
location of the nucleus and can therefore vary widely in magnitude.
For the purposes of our statistical analysis here we shall take an even more
radical point of view and choose the $A_{i}^{(\alpha)}$ at random. To be
specific, we will choose the $A_{i}^{(\alpha)}$ from a uniform distribution
within the interval $[0,1]$ and normalize them afterwards according to
$\sum_{i}A_{i}^{(\alpha)}=1$ for each central spin. The data to be presented below
is obtained by averaging over typically 500 random realizations of 
coupling parameters. Note that the Hamiltonian matrix represented
in the usual basis of tensor-product eigentstates of $S_{\alpha}^{z}$,
$I_{i}^{z}$ is always real and symmetric. Therefore, the natural candidate
for a random matrix description of such systems is the Gaussian
orthogonal ensemble (GOE) \cite{Guhr98}.

In the important case of a single central spin, $N_{c}=1$, the above model
has the strong mathematical property of being integrable
\cite{Gaudin76,Bortz07}. Moreover, this integrability is 
particularly robust as it is independent of the choice of the
coupling parameters $A_{i}^{(1)}$ and the length of the spins which can even be 
chosen individually \cite{Gaudin76,Bortz07}. In fact, the model (\ref{genmod}) 
for a single central spin has been the basis of numerous theoretical studies on
decoherence properties of quantum dot spin qubits; see, for example,
Refs.~\cite{Khaetskii02,Merkulov02,Schliemann02,Coish08,Cywinski09}, for
reviews also \cite{Schliemann03,Zhang07,Klauser07}. 
It is an interesting question,
both from a practical as well as from an abstract point of view, 
to what extend the results of these investigations are linked to the 
integrability of the underlying idealized model. In particular, what
changes may occur if the Hamiltonian deviates from the above simple case
$N_{c}=1$ by, e.g., involving more than one central spin?
Previous investigations of decoherence properties, 
making strongly restrictive assumptions
on the coupling constants,
predicted a significant dependence on whether
the number of central spins is even or odd \cite{Dobrovitski03,Melikidze04}.
In the following we will investigate Hamiltonians of the general type
(\ref{genmod})
within the framework of level statistics, i.e. generic
spectral characteristics \cite{Guhr98}. For other studies of
interacting quantum many-body systems using this method see e.g.
Refs.~\cite{Montambaux93,Georgeot98,Avishai02}.

\begin{figure}
   \centering
   \includegraphics[height=10cm,keepaspectratio=true]{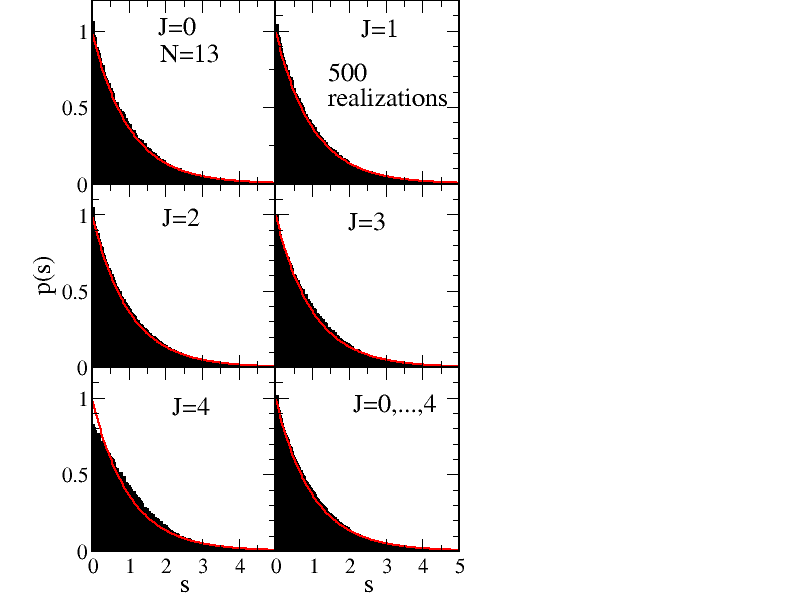}
   \caption{(Color online) 
Nearest-neighbor level spacing for a system of a single central spin
$S_{1}=1/2$ and 13 bath spins of length $I=1/2$. The red curve is the 
exponential $p(s)=e^{-s}$.}
   \label{fig1}
\end{figure}

\begin{figure}
   \centering
   \includegraphics[height=10cm,keepaspectratio=true]{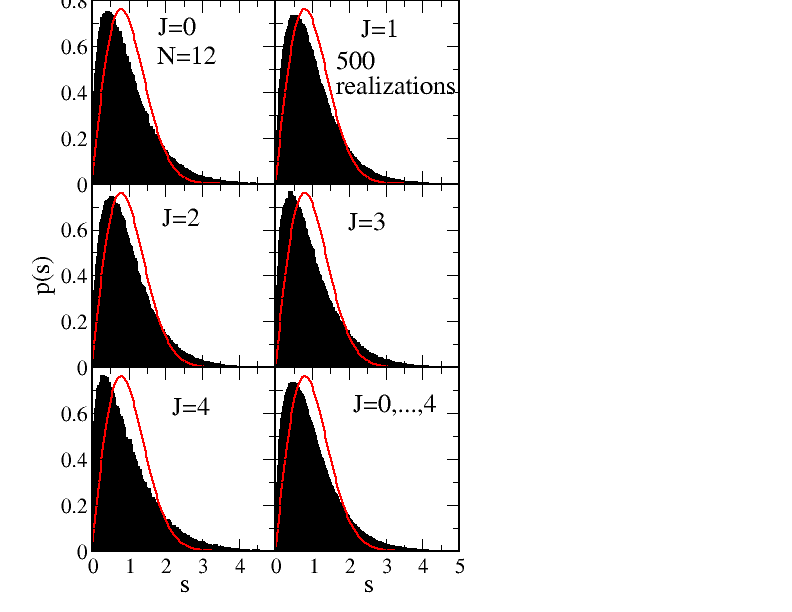}
   \caption{(Color online) 
Nearest-neighbor level spacing for a system of a two central spins
$S_{1}=S_{2}=1/2$ and 12 bath spins of length $I=1/2$. The red curve is the 
GOE Wigner surmise.}
   \label{fig2}
\end{figure}

The spectra generated numerically
from the Hamiltonian (\ref{genmod}) clearly have a nontrivial overall
structure, i.e. the locally averaged density of states is not constant
as a function of energy \cite{Schliemann02,Schliemann03}. 
Therefore an unfolding of these spectra has to be
performed which results in a transformation onto a new spectral variable $s$
such that the mean level density is equal to unity \cite{Guhr98}.
We have compared several standard numerical unfolding procedures and made sure 
that they yield consistent results.
Fig.~\ref{fig1} shows the probability distribution $p(s)$
for the nearest-neighbor level spacing for a system of a single central spin
$S_{1}=1/2$ and 13 bath spins of length $I=1/2$ for several subspaces
of the total angular momentum $\vec J=\vec S_{1}+\sum_{i}\vec I_{i}$ where
each multiplet is counted as a single enery level.
The subspaces of highest $J=5,6,7$ have been discarded, and in the
bottom right panel all probability distributions are joined.  
As to be expected for an integrable model, the level statistics follow
a Poisson distribution resulting in an exponential 
level spacing distribution $p(s)=e^{-s}$. This is in contrast to the case
of two central spins $S_{1}=S_{2}=1/2$ shown in Fig.~\ref{fig2}. 
Here level repulsion takes clearly place, $p(0)=0$, 
although the data considerably
deviates from the Wigner surmise for the GOE \cite{Guhr98},
$p(s)=(\pi/2)s\exp(-(\pi/4)s^{2})$. 

\begin{figure}
   \centering
   \includegraphics[height=10cm,keepaspectratio=true]{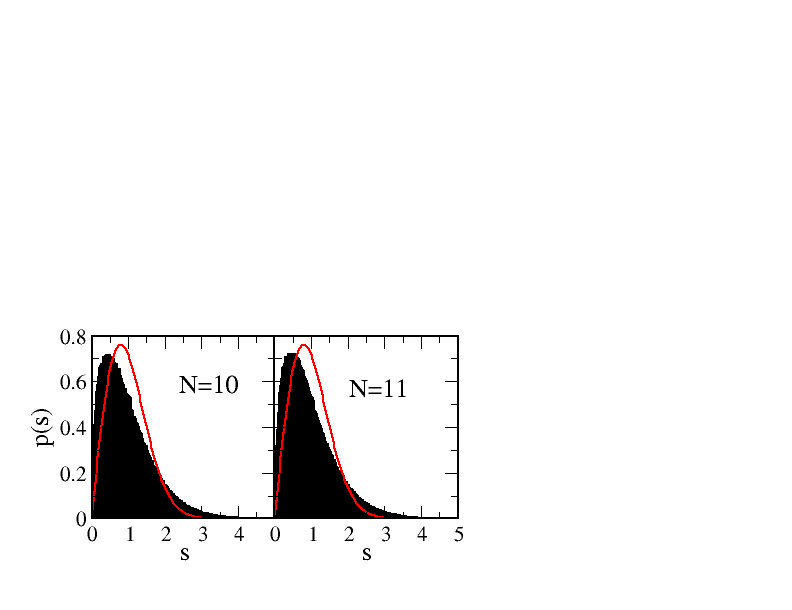}
   \caption{(Color online)
The same data as in the bottom right panel of
Fig.~\ref{fig2} but for $N=10$ and $N=11$ bath spins.}
   \label{fig3}
\end{figure}

Obviously, our numerical studies are technically restricted to rather small
system sizes, $N_{c}+N\leq 14$. This limitation, however, does not affect
our results for the level statistics as demonstrated in Fig.~\ref{fig3}
where we have plotted the same data as in the bottom right panel of
Fig.~\ref{fig2} but for $N=10$ and $N=11$ bath spins.
This insensitivity to the system size
seen in the figure is a natural consequence of the unfolding of the spectra.

\begin{figure}
   \centering
   \includegraphics[height=10cm,keepaspectratio=true]{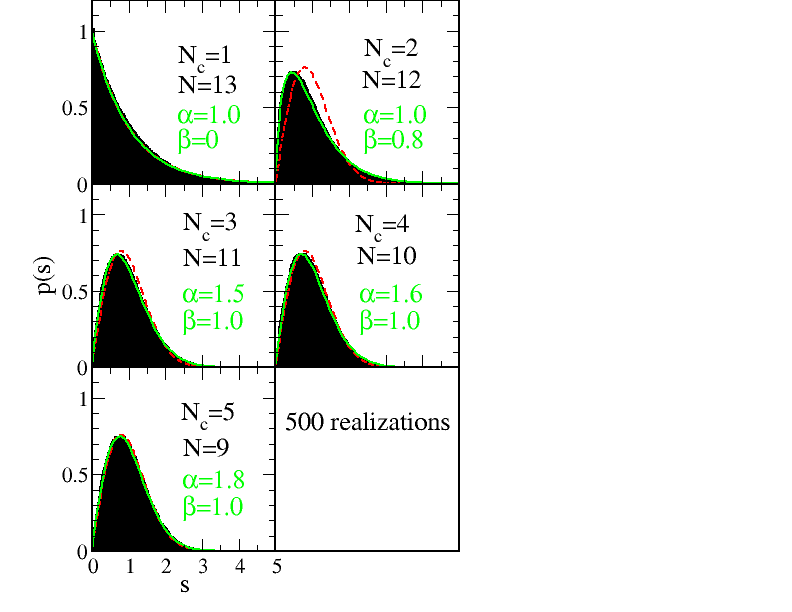}
   \caption{(Color online)
Joint level spacing distribution for
$J=0,\dots,4$ and increasing number of central spins. The red dashed lines are
the exponential function ($N_{c}=1$) and the Wigner surmise ($N_{c}>1$). The
green solid lines are a fit to the generalized Brody distribution 
(\ref{ansatz}).}
   \label{fig4}
\end{figure}

Fig~\ref{fig4} shows the joint level spacing distribution for
$J=0,\dots,4$ and increasing number of central spins, where
$p(s)$ approaches closer and closer the Wigner surmise. To quantify this
observation we use the ansatz
\begin{equation}
p(s)=Bs^{\beta}e^{-As^\alpha}
\label{ansatz}
\end{equation}
with 
\begin{eqnarray}
B & = & \alpha(\Gamma((\beta+2)/\alpha))^{\beta+1}
/(\Gamma((\beta+1)/\alpha))^{\beta+2}\,,\\
A & = & \left(\Gamma((\beta+2)/\alpha)/\Gamma((\beta+1)/\alpha)\right)^{\alpha}
\end{eqnarray}
such that $\int p(s)=\int sp(s)=1$. Clearly, $\alpha=1$, $\beta=0$
corresponds to an exponential distribution whereas $\alpha=2$, $\beta=1$
reporduces the Wigner surmise. The above ansatz generalizes the Brody 
distribution given by $\alpha=\beta+1$ \cite{Brody73,Guhr98}.
As seen in Fig.~\ref{fig4}, the numerical data is very well described by the
above distribution, and for the case of $5$ central spins the Wigner surmise
of the GOE is almost reached.
In particular, our level statistics do not show any 
odd/even-effects with respect to the number of central spins
as predicted in Refs.~\cite{Dobrovitski03,Melikidze04}. 
We attribute this difference to the 
strongly restrictive assumptions made there
giving rise to additional symmetries.

Moreover, in the case of two central spins it is instructive to rewrite
the Hamiltonian in the following form,
\begin{eqnarray}
{\cal H} & = & 
(\vec S_{1}+\vec S_{2})\cdot
\sum_{i}\frac{1}{2}(A_{i}^{(1)}+A_{i}^{(2)})\vec I_{i}\nonumber\\
 & + & (\vec S_{1}-\vec S_{2})\cdot
\sum_{i}\frac{1}{2}(A_{i}^{(1)}-A_{i}^{(2)})\vec I_{i}\,.
\label{two}
\end{eqnarray}
The two central spins $\vec S=\vec S_{1}+\vec S_{2}$ can couple
to $S=0,1$. Since the coupling to the singlet $S=0$ vanishes, 
the first line in Eq.~(\ref{two}) is just the integrable Hamiltonian
of a single central spin $S=1$, whereas the second line can be viewed as a 
perturbation. This term vanishes if the coupling constants are still random but
chosen to be the
same for each spin, $A_{i}^{(1)}=A_{i}^{(2)}$, resulting in an integrable
model of two central spins, a prediction we have explicitly verified in our
numerics; the latter model was also studied numerically in 
Ref.~\cite{Dobrovitski03}.

The models studied so far have a {\em common} spin bath, i.e. each bath spin
couples without any further restriction to each central spin. Regarding
the generic example of two neighboring quantum dot spin qubits this is not
particularly realistic since in this geometry one can obviously identify groups
of nuclear spins which couple strongly to one of the electron spins but
weakly to the other. The extreme case is given by two {\em separate} spin baths
where the central spins can be coupled via an exchange interaction
\cite{Burkard99,Schliemann01},
${\cal H}^{\prime}={\cal H}+J_{ex}\vec S_{1}\cdot\vec S_{2}$.
Here we find numerically that even arbitrary small exchange parameters
$J_{ex}$ break integrability and lead to level repulsion. The corresponding level
spacing distributions, however, are less accurately described by the ansatz
(\ref{ansatz}). On the other hand, for large $|J_{ex}|$ the system approaches 
the integrable scenario since then the singlet and triplet subspace
of the central spins are energetically more and more separated.

Let us now discuss the influence of an external magnetic field coupling
to the central spins.
In the case $N_{c}=1$ the resulting model is known to be
integrable \cite{Gaudin76,Bortz07}, 
and also for $N_{c}>1$ the Hamiltonian can still be 
represented as a real and symmetric matrix. Indeed, we have not seen
any qualitative difference in the level spacing distribution with and
without an external magnetic field. In particular, we have not found any 
sign for a transition between the Gaussian orthogonal to the unitary
ensemble (as appropriate for systems lacking time reversal symmetry
\cite{Guhr98}). In this sense, the application of an external magnetic field
can be viewed as a ``false symmetry breaking'' which still preserves
a ``non-conventional time-reversal invariance'' \cite{Avishai02,Haake00}.
We note that recent theoretical works predict different time 
dependencies of spin dynamics in different magnetic-field regimes
\cite{Coish08,Cywinski09}.
These observations are not reflected by the level statistics. Thus,
decoherence and the occurrence of integrability or chaoticity are independent 
phenomena in such systems, at least as far as the role of magnetic fields is
concerned.

The data presented so far was obtained for bath spins of length $I=1/2$.
Motivated by the large nuclear spins in semiconductor materials, we
have also performed simulations for $I=1$ which also do not show any 
qualitative difference to the previous case. This is indeed to be expected
since a spin bath of $I=1$ can be obtained from a bath with
$I=1/2$ and twice the number of bath spins by grouping the spins
into pairs and chosing the coupling parameters to be the same in each pair.
Similar considerations apply to higher bath spins.

\begin{figure}
   \centering
   \includegraphics[height=7cm,keepaspectratio=true]{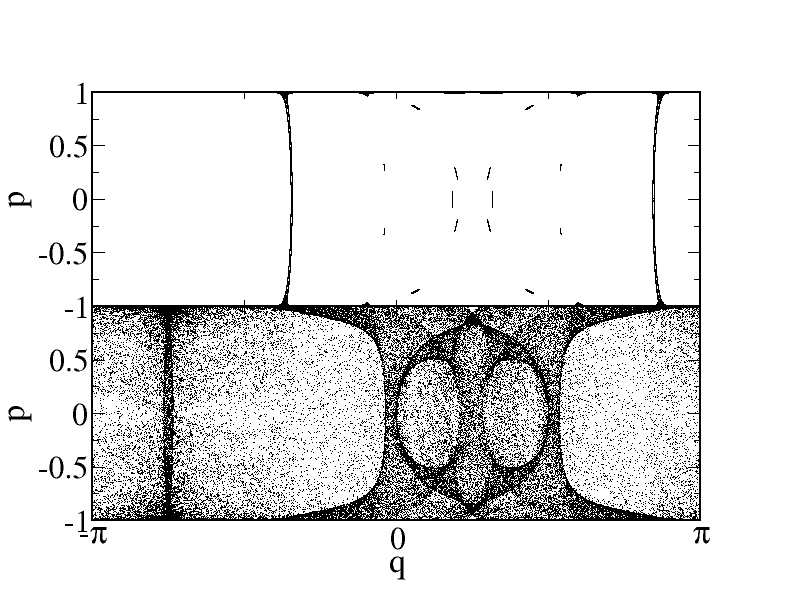}
   \caption{A cut through the plane $I^{z}_{1}=I^{z}_{2}=0$ of two different
phase space orbits demonstrating the close vicinity of regular and
chaotic dynamics in a system of two central spins (see text).}
   \label{fig5}
\end{figure}

Let us come back to the case of two central spins. As seen in 
Figs.~\ref{fig2},\ref{fig4}, this system appears to lie in between
the integrable case and the predictions of random matrix theory.
Thus, in the light of the Bohigas-Giannoni-Schmitt conjecture
\cite{Bohigas84}, it is natural to speculate that the classical 
counterpart of this
system has a mixed phase space consisting of areas of regular and of 
chaotic dynamics.
The classical limit of a quantum spin system is naturally obtained via
spin-coherent states, and a pair of classical canonically conjugate
variables $p$, $q$ for each spin is given by $p=\cos\vartheta$, $q=\varphi$,
where $\vartheta$, $\varphi$ are the usual angular coordinates of
the classical spin unit vector \cite{Schliemann98}.
We have performed numerical Runge-Kutta simulations of such classical dynamics
for $N_{c}=2$ where one can easily treat systems of several thousand bath spins.
However, to avoid the complications of such a high-dimensional phase space
let us concentrate on the smallest nontrivial case of just two bath spins.
Here we find indeed a close vicinity of regular and chaotic dynamics.
An example is shown in Fig.~\ref{fig5} where we have plotted in the
top panel a cut through the plane $I^{z}_{1}=I^{z}_{2}=0$
as a function of $p=S^{z}_{1}$, $q=\tan^{-1}(S^{y}_{1}/S^{x}_{1})$. The coupling
constants are given by $A_{1}^{(1)}=1-A_{2}^{(1)}=0.75$,
$A_{1}^{(2)}=1-A_{2}^{(2)}=0.52$, and the initial condition is
$\vec S_{1}=-\vec S_{2}=(0,0,-1)$, $\vec I_{1}=(1,0,0)$, $\vec I_{2}=(0,1,0)$.
This arrangement leads obviously to very regular dynamics, in stark contrast
with the bottom panel where we have used the same initial condition but
introduced a minute change in one pair of coupling constants, 
$A_{1}^{(2)}=1-A_{2}^{(2)}=0.5195$, resulting in a clearly chaotic orbit with an
inhomogeneous phase space filling.
Note that the observation that certain phase space curves are overlaid
in the figure is due to the fact that the remaining six phase space variables
are not uniquely determined by the condition $I^{z}_{1}=I^{z}_{2}=0$
and the conserved quantities ${\cal H}=0$, $\vec J=(1,1,0)$ but occur in 
several branches.

In summary, we have investigated central spin models via
nearest-neighbor level spacing distributions. As the number of 
central spins increases a transition from Poissonian statistics to the 
Gaussian orthogonal ensemble sets in which can be described by a generalized 
Brody
distribution. These observations are not affected by the finite system size in
our numerical simulations and are 
unaltered upon applying an external 
magnetic field. In the transition region, the classical counterparts of the
models studied have mixed phase space.

I thank M. Brack, K. Richter, S. Schierenberg, and T. Wettig for
useful discussions. This work was supported by DFG via SFB 631.

\end{document}